\documentclass[10pt,pre,a4paper,twocolumn,superscriptaddress,showpacs]{revtex4-1}
\usepackage{amsmath}
\usepackage{amssymb}
\usepackage{amsthm}
\usepackage{mathtools}
\usepackage{graphicx}
\usepackage{grffile}
\usepackage{color}
\usepackage[dvipsnames]{xcolor}

\newcommand{\xx}[1]{{#1}}

\begin{document}

\author{Claus Heussinger} \affiliation{Institute for theoretical
  physics, Georg August University Göttingen, Friedrich Hund Platz 1,
  37077 Göttingen, Germany} \title{Start-up shear of spherocylinder
  packings: effect of friction}

\begin{abstract}

  We study the response to shear deformations of packings of long
  spherocylindrical particles that interact via frictional forces with
  friction coefficient $\mu$. The packings are produced and deformed
  with the help of molecular dynamics simulations combined with
  minimization techniques performed on a GPU.  We calculate the linear
  shear modulus $g_\infty$, which is orders of magnitude larger than
  the modulus $g_0$ in the corresponding frictionless system.  The
  motion of the particles responsible for these large frictional
  forces is governed by and increases with the length $\ell$ of the
  spherocylinders. One consequence of this motion is that the shear
  modulus $g_\infty$ approaches a finite value in the limit
  $\ell\to\infty$, even though the density of the packings vanishes,
  $\rho\propto\ell^{-2}$.
  By way of contrast, the frictionless modulus decreases to zero,
  $g_0\sim\ell^{-2}$, in accordance with the behavior of density.
  Increasing the strain beyond a value $\gamma_c\sim \mu$, the packing
  undergoes a ``shear-thinning'' transition from the large frictional
  to the smaller frictionless modulus when contacts saturate at the
  Coulomb inequality and start to slide. In this regime, sliding
  friction contributes a ``yield stress'' $\sigma_y=g_\infty\gamma_c$
  and the stress behaves as $\sigma=\sigma_y+g_0\gamma$. The interplay
  between static and sliding friction gives rise to hysteresis in
  oscillatory shear simulations.

\end{abstract}

\maketitle

\section{Introduction}

In this work we deal with the linear elastic properties of packings of
long rod-like particles. Assemblies of this sort occur in a variety of
different systems from wool fibers in felt and other textiles, rods in
filter applications, or as reinforcement, to micro- and nano-sized
systems, like fd-virus colonies or semiflexible biopolymers in tissue
and the cytoskeleton. Aspect ratios and interaction forces are
manifold. In macroscopic systems, frictional forces are essential
(e.g. to hold textiles together). In cytoskeletal systems one has
Brownian forces and protein-mediated adhesion (bonding) between the
fibers.

\begin{figure}[t]
  \includegraphics[width=0.23\textwidth]{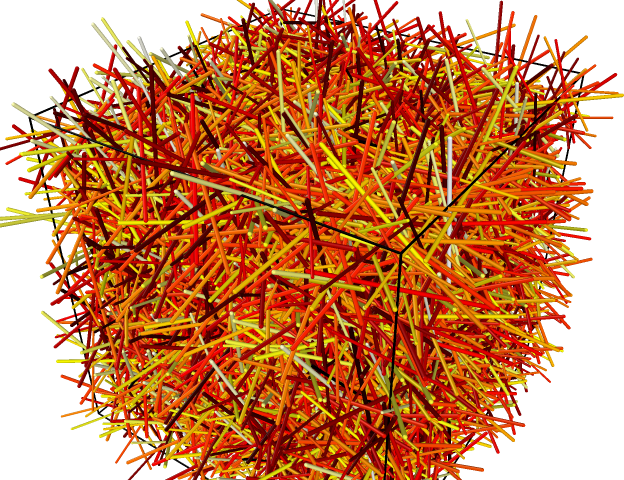}
  \includegraphics[width=0.23\textwidth]{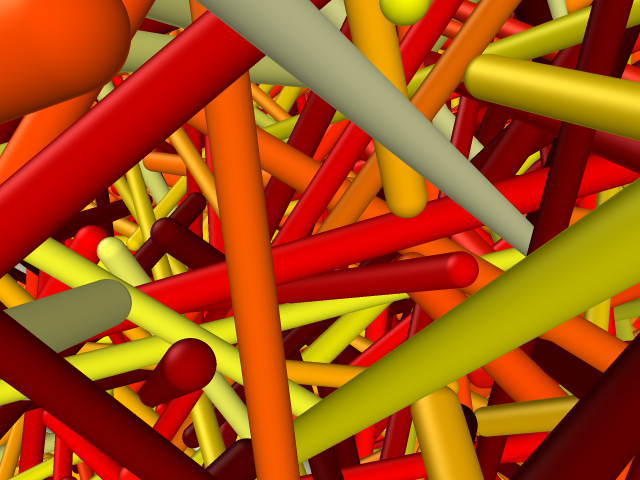}\vspace{2mm}
  \includegraphics[width=0.40\textwidth]{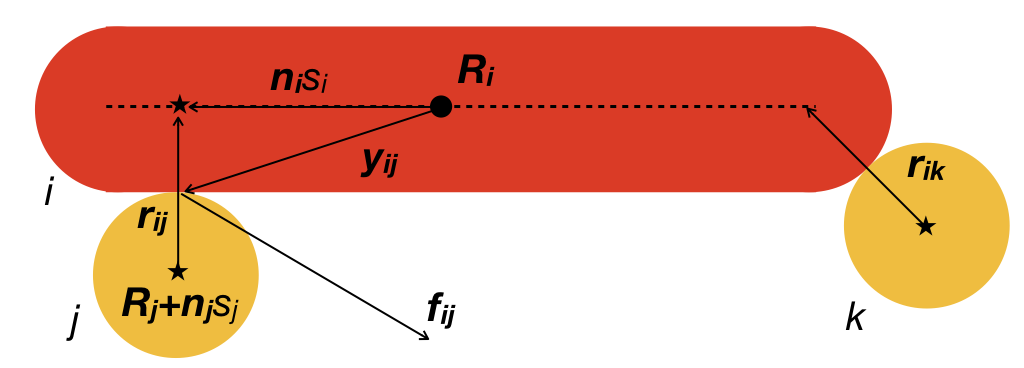}
  \caption{Top: packing of 6144 spherocylinders with aspect ratio
    $\alpha=40$. Volume fraction $\phi=0.1$, i.e. $90\%$ of space is
    empty. Zoom into packing illustrates these large voids. Color
    signals orientation. Pictures prepared with OVITO. \xx{Bottom:
      sketch illustrating the interaction between
      spherocylinders. Particle $i$ is drawn as projection into the
      drawing plane, while particles $j$ and $k$ are oriented
      perpendicular to this plane. Therefore, they appear as
      circles. Particle $j$ is a side contact to $i$, while particle
      $k$ illustrates the occurence of an end contact.}}
  \label{fig:network}
\end{figure}

Here, we are interested in the effects of frictional forces on the
elastic shear modulus of a packing of non-Brownian spherocylindrical
particles (see Fig.~\ref{fig:network}). A spherocylinder \xx{(SC)}
consists of a cylinder and two hemispherical caps at the two ends. In
a previous publication~\cite{heussinger20:_packin} we have dealt with
the same system, but in the absence of friction. Steric hindrance
plays a crucial role for the motion of particles, which are tightly
caged by their neighbors. Motion along the long cylinder axis,
however, is not restrained by the surrounding. This motion induces
sliding of the contacts on the surface of the cylinders and is
expected to give rise to large frictional forces. We therefore expect
friction to modify the shear modulus dramatically.

Packings of non-spherical particles have received considerable
attention in recent years. In particular the question of maximally
dense packings has been the subject of works on particles of various
shapes~\cite{RevModPhys.82.2633}. Most of these particles are rather
compact and more or less sphere-like, quite different from the long,
thin rods discussed in this contribution. \xx{Increasing the length of
  the rods from zero, the density reaches a maximum
  \cite{zhao12:_dense,willi03} before dropping steadily. Assembled in
  random fashion, long rods make rather dilute packings, with the
  density decreasing with particle length as $\rho\propto
  \ell^{-2}$~\cite{phi96}.}  In terms of particle volume fraction
$\phi=\rho V_{\rm sc}$ the dependence is $\phi\propto \ell^{-1}$. This
is a consequence of the different scaling of particle volume $V_{\rm
  sc}$ and excluded volume~\cite{onsager49} with diameter $d$ and
length $\ell$ of rod-like particles
\begin{eqnarray}\label{eq:v_sc}
 V_{\rm sc} \sim d^2\ell &\qquad& V_{\rm excl} \sim d\ell^2\,.
\end{eqnarray}
Neglecting correlations between particles (random contact model,
Ref. \cite{phi96}) the average number of contacts of a particle is
\begin{eqnarray}\label{eq:z_rcp}
 z \sim N\frac{V_{\rm excl}}{V} &=& \phi\frac{V_{\rm excl}}{V_{\rm sc}} \,.
\end{eqnarray}
With the number of contacts fixed at the jamming threshold $z_c$ one
gets for the jamming density
\begin{equation}\label{eq:phi_rcp}
  \phi \sim \frac{z_cd}{\ell}\,.
\end{equation}
Measured values for $z_c$ range from approximately 8 to 10
\cite{rod05,willi03,phi96,Blouwolff_2006}. The latter value represents
the classical Maxwell counting \cite{calladine78} for particles with
one rotational symmetry. The reduced value of 8 results when
translations along the long axis are also regarded as a
symmetry~\cite{heussinger20:_packin}. In that paper it was shown that
these two limiting cases can be combined by a more general counting
procedure that accounts for the fraction $f$ of spherocylinders with
end contacts (both ends need to be constrained). This can be written
in analogy to Eq.~(\ref{eq:z_rcp}) as
\begin{eqnarray}\label{eq:f_end}
 f \sim \left(N\frac{V_{\rm sc}}{V}\right)^2 \sim \phi^2\,.
\end{eqnarray}
End contacts break the translational symmetry such that the jamming
threshold in terms of the contacts is $z_c= 8 +2f$.

\xx{Previous work on packings of fibers and rods is mainly
  computational. Statistical properties of packings of rods are
  calculated in \cite{willi03,zhao12:_dense,wouterse09}. Meng et al
  \cite{MENG2016176} highlight the dependence on the amount of order
  in the packing, while \cite{Pournin2005} show that order may develop
  in response to repeated tapping. Bending flexibility is introduced
  either within bead-spring models \cite{rod05,PhysRevLett.118.068002}
  or by coupling rigid rods together
  \cite{langston15:_discr,PhysRevE.80.016115}.}

\xx{Experimental approaches to measure density in rod packings are
presented in \cite{parkhouse95,phi96}, Ref. \cite{Blouwolff_2006} even
manage to determine the number of contacts. Ekman et
al. \cite{PhysRevLett.113.268001} highlight a correlation effect that
goes beyond the random contact model of Ref. \cite{phi96} and that
shows up in the distribution of contacts.}

\xx{Going beyond static packings rheological properties are discussed in a
variety of contexts. Some computational approaches are reviewed in
\cite{doi:10.1146/annurev-fluid-122316-045144}. Steady shear in
systems of short spherocylinders has been studied in Nath {\it et
  al.}~\cite{nath19:_rheol}. Key finding was that frictional
interaction forces supress alignment during shear. Similar systems
have been studied in three \cite{PhysRevE.96.062903,mahajan19:_fluid}
and two spatial dimensions
\cite{PhysRevE.101.032907,PhysRevE.100.032906,PhysRevE.81.051304}. Experimental
work frequently considers the interplay of (frictional) contact and
hydrodynamic forces in suspensions
\cite{tapia_shaikh_butler_pouliquen_guazzelli_2017,egres05,PhysRevE.84.031408,PhysRevFluids.3.074301}.}

\xx{Here, we are concerned with the response of rod packings to small
deformations, within the linear regime. Particular emphasis is put on
the relevance of frictional forces. A related study was presented in
Ref.~\cite{PhysRevE.80.016115} that deals with compressed packings of
elastic fibers that can bend and stretch. It turns out that packings
without friction have a negligible shear modulus as compared to
frictional packings. In the following, we will present a similar
phenomenon in packings of rigid rods. By carefully evaluating the
dependence on rod length, we find the origin of the large frictional
modulus, and explain it in terms of the dominant deformation modes.
}

\section{Model}

We study three-dimensional (3d) packings of spherocylindrical
particles $i=1\ldots N$ of length $\ell_i$ and diameter $d_i$. The
particles interact via repulsive contact forces similar to those from
models for spheres. A contact between particles $i$ and $j$ is
established whenever the shortest distance between the backbones,
$r_{ij} = |\mathbf{r}_{ij}|$, is less than their average diameter
$d_{ij}=(d_{i}+d_{j})/2$.

The distance vector can be written as
\begin{equation}\label{eq:Rij}
  r_{ij}\mathbf{\hat e}_{ij} = \mathbf{R}_i+\mathbf{\hat n}_is_i -
  (\mathbf{R}_j+\mathbf{\hat n}_js_j)\,,
\end{equation}
where $\mathbf{R}_i$ is the position of the center of mass of particle
$i$, $\mathbf{\hat n}_i$ represents the direction of the particle
backbone, and $s_i\in[-\ell_i/2,\ell_i/2]$ is the arclength parameter
along the backbone that specifies where the shortest distance between
$i$ and $j$ is reached. By definition the direction of the contact
$\mathbf{\hat e}_{ij}$ is perpendicular to both backbones
($\mathbf{\hat e}_{ij}\cdot \mathbf{\hat n}_{i/j}=0$), except for
cases where the shortest distance is reached at an end of one or both
of the SC (i.e.  $s_i=\pm \ell_i/2$). The actual force is applied halfway
along the vector $\mathbf{r}_{ij}$ at $\mathbf{y}_{ij} = \mathbf{\hat
  n}_is_i+\mathbf{r}_{ij}/2$ away from the center of mass. (with a
small correction for unequal-sized particles). This is, in general,
very close to the surface of the two particles.

The force $\mathbf{f}_{ij}$ on particle $i$ from the contact with $j$
has components normal $\mathbf{f}^n_{ij}$ and tangential
$\mathbf{f}^t_{ij}$ to the particle surface. They are calculated as
in the Cundall-Strack model \cite{cundall79}
\begin{align}\label{eq:contactforce_definition}
  \mathbf{f}^n_{ij}&=[-k_n \delta_{ij} \mathbf{\hat e}_{ij} - c_n\mathbf{v}^n_{ij}] ,\\
  \mathbf{f}^t_{ij}&=[-k_t \boldsymbol{\xi} ^t_{ij} - c_t\mathbf{v}^t_{ij}].\nonumber
\end{align}

Here, the normal direction $\mathbf{\hat
  e}_{ij}=\mathbf{r}_{ij}/r_{ij}$ points from particle $j$ to $i$ at
the point of application of the force. The normal overlap
$\delta_{ij}= d_{ij}-r_{ij}$ is a positive quantity. The tangential
overlap $\boldsymbol{\xi} ^t_{ij}$ is the displacement tangential to
the surface of the SC, which accumulates during the lifetime of the
contact. The relative velocity $\mathbf{v}_{ij}^{\rm con}$ at the
contact is split into normal $\mathbf{v}^n_{ij}$ and tangential
components $\mathbf{v}^t_{ij}$. It derives from the center of mass
translational $\mathbf{v}_i$ and rotational motion
$\boldsymbol\omega_i$ as $\mathbf{v}_{ij}^{\rm con} = (\mathbf{v}_{i}
- \mathbf{v}_{j}) -
\mathbf{y}_{ij}\times\boldsymbol\omega_i+\mathbf{y}_{ji}\times\boldsymbol\omega_j$.

The parameters $k_n$ and $k_t$ are spring constants, $c_n$ and $c_t$
are viscous damping constants. Solid sliding friction is taken into
account replacing $\mathbf{f}^t$ by $\mu|\mathbf{f}^n|(\mathbf{f}^t /
|\mathbf{f}^t|)$, whenever the Coulomb inequality
\begin{equation}\label{eq:coulomb}
  |\mathbf{f}^t|<\mu |\mathbf{f}^n|\,,
\end{equation}
is violated.
We also consider the frictionless limit $\mu=0$, in which case we
still keep the tangential viscous force $\propto c_t$.

The equations of motion for particle $i$ are
\begin{equation}
 m \mathbf{\ddot R}_i=\sum_j \mathbf{f}_{ij}
\end{equation}
\begin{equation}
 \mathbf{I}_i\cdot \boldsymbol{\dot{\omega}}_i=\sum_j \mathbf{y}_{ij} \times \mathbf{f}_{ij}
\end{equation}
where $\mathbf{I}_i$ is the moment of inertia of particle $i$
calculated for a spherocylinder with a homogeneous mass density.

We have set $k_n=1$, $k_t/k_n=2/7$, $c_n=0.5$ and $c_t/c_n=0.1$.
\xx{These values are standard choices. In particular $c_n$ is chosen
  such that damping is sufficiently strong without leading to
  overdamped motion. For truly elastic response, the value of the
  dissipative coefficients do not matter.}  All particles have the
same mass $m=1$ and aspect ratio $\alpha=\ell/d$. Half of the
particles have $d=1$, the other half have $d=1.4$. System sizes are
chosen such that the linear dimension of the simulation box is at
least three times the length of the simulated SC.  \xx{The unit of
  energy density (stress or modulus) is $k_n/d$.} Times are expressed
in units of the elastic collision time $\sqrt{m/k_n}$.

\section{Results: response to shear}

\subsection{Preparing packings for shear deformation}

A stable, i.e. force-equilibrated, packing of SCs of a given aspect
ratio (see Fig.~\ref{fig:network}) is produced with the help of the
FIRE minimization~\cite{PhysRevLett.97.170201}. During this initial
minimization no dissipative nor frictional forces are present,
i.e. $k_t=c_t=c_n=0$. The response of such frictionless packings to
quasistatic shear has been studied in
Ref.~\cite{heussinger20:_packin}. Here we are interested in how
frictional forces modify the response. Thus, after minimization,
frictional forces with parameters specified above are turned on, while
the simulation box is deformed at a constant strainrate
$\dot\gamma=10^{-7}$. Lees-Edwards boundary
conditions~\cite{lee_edward} are used here. The strainrate needs to be
chosen small enough to reproduce the quasistatic results at zero
friction, $\mu=0$. We have checked that this is the case, see
e.g. Fig.~\ref{fig:stress_varmu}. For the linear response properties
only small strains $\Delta\gamma\approx 10^{-4}$ are needed.

\subsection{Large friction limit: $\mu\to\infty$}

If $\mu$ is infinite, the Coulomb inequality, Eq.~(\ref{eq:coulomb}),
is ineffective and provides no restriction on the frictional forces
$\mathbf{f}^t$. A frictional interaction force at a contact then acts
similar to a conservative force from an elastic bond with spring
constant $k_t$ (neglecting a small effect from $c_t$). The magnitude
of the tangential displacement $\boldsymbol{\xi}^t$ then represents
the extension of this spring. This limit is achieved when the response
does not change anymore with $\mu$.  Here we use $\mu=10$.

\begin{figure}[ht]
  \includegraphics[width=0.4\textwidth]{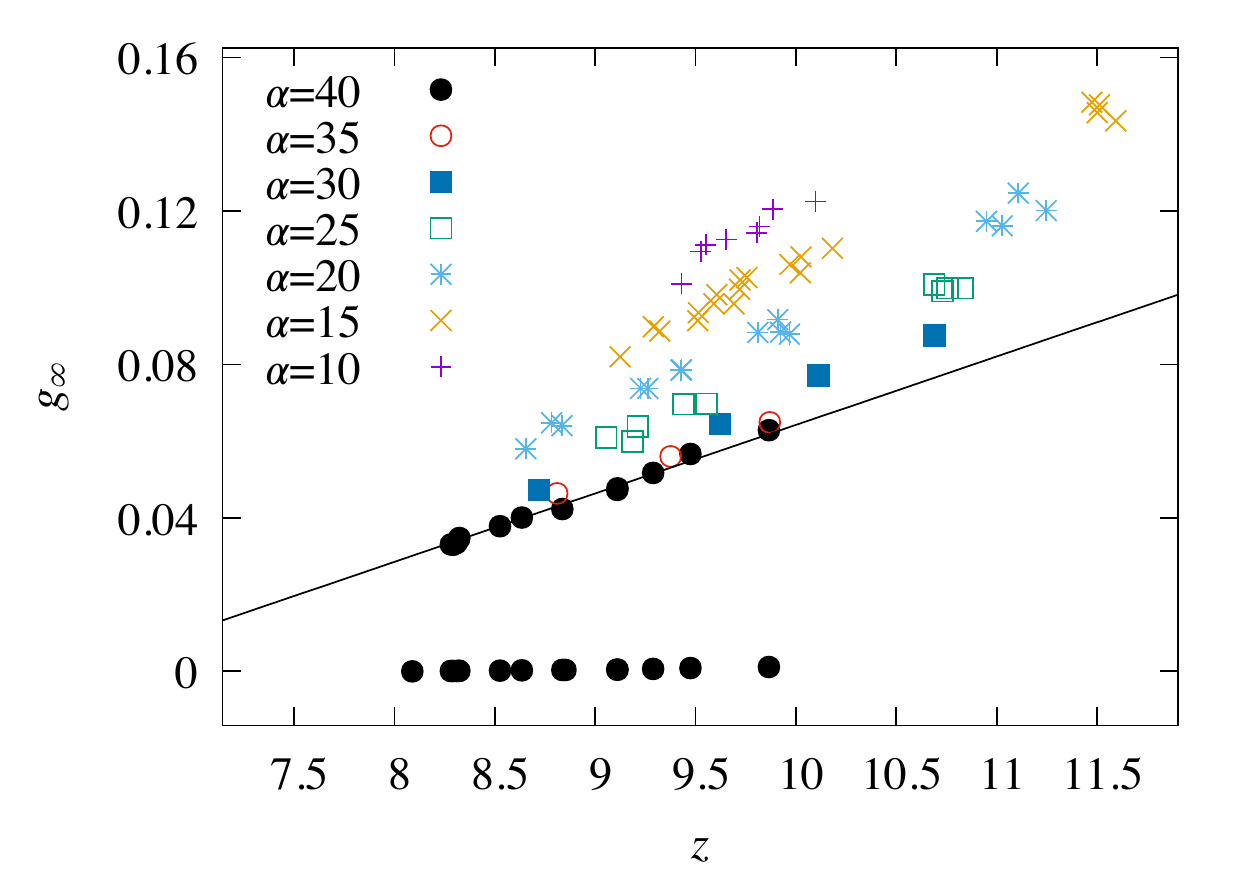}
   \caption{Shear modulus $g_\infty$ vs. contacts $z$ for infinite friction
    $\mu\to\infty$ and various aspect ratios. Line is a fit $g\sim
    (z-z_0)$ to the data with $\alpha=40$. The data points at nearly
    zero correspond to the frictionless limit ($\alpha=40$) where the
    modulus is on the order of $10^{-4}$.}
  \label{fig:modulus_z}
\end{figure}

The stress is calculated from the virial expression
\cite{PhysRevE.100.032906,C001085E}
\begin{equation}\label{eq:stress_def}
  \sigma = \frac{1}{V}\sum_{k<l} f^x_{kl}R_{kl}^y =
  \frac{Nz}{2V}\langle f^xR^y\rangle_c
\end{equation}
where the latter expression denotes an average over all contacts $N_c$
with $z=2N_c/N$ the average number of contacts per SC. \xx{Due to the
  relation $\mathbf{R_{kl}} = \mathbf{R_{k}} -\mathbf{R_{l}} =
  \mathbf{y_{kl}} - \mathbf{y_{lk}}$ one may also write the stress in
  terms of the lever arms as $\sigma = \frac{1}{V}\sum_{k,l}
  f^x_{kl}y_{kl}^y$.} From the stress-strain relation $\sigma(\gamma)$
at small strains the linear shear modulus is calculated as the slope,
$g=d\sigma/d\gamma$. In Fig.~\ref{fig:modulus_z},
$g_\infty=g(\mu\to\infty)$ is plotted for various configurations with
different spherocylinder aspect ratio $\alpha=\ell/d$ and volume
fraction $\phi$. As the control parameter the contact number $z$ is
used.

\xx{The moduli can be fit to the form $g\sim z-z_0$ with $z_0$ ranging
between $6$ and $6.5$ depending on the aspect ratio. However, it is
also possible to fit all different $\alpha$ with one $z_0\approx
6.5$. This threshold value is larger than the frictional jamming limit
$z_J=4$. As our packings are produced by energy minimization in the
absence of friction, they necessarily have a contact number $z>8$. As
soon as the coordination drops markedly below this value, energy
minimization takes the packing to $z\to0$. Thus, the threshold $z_0$
cannot be reached by using our protocol. It might be interesting to
study different packing-generation protocols in order to reach to
lower $z$.}

\begin{figure*}[ht]
  \includegraphics[width=0.32\textwidth]{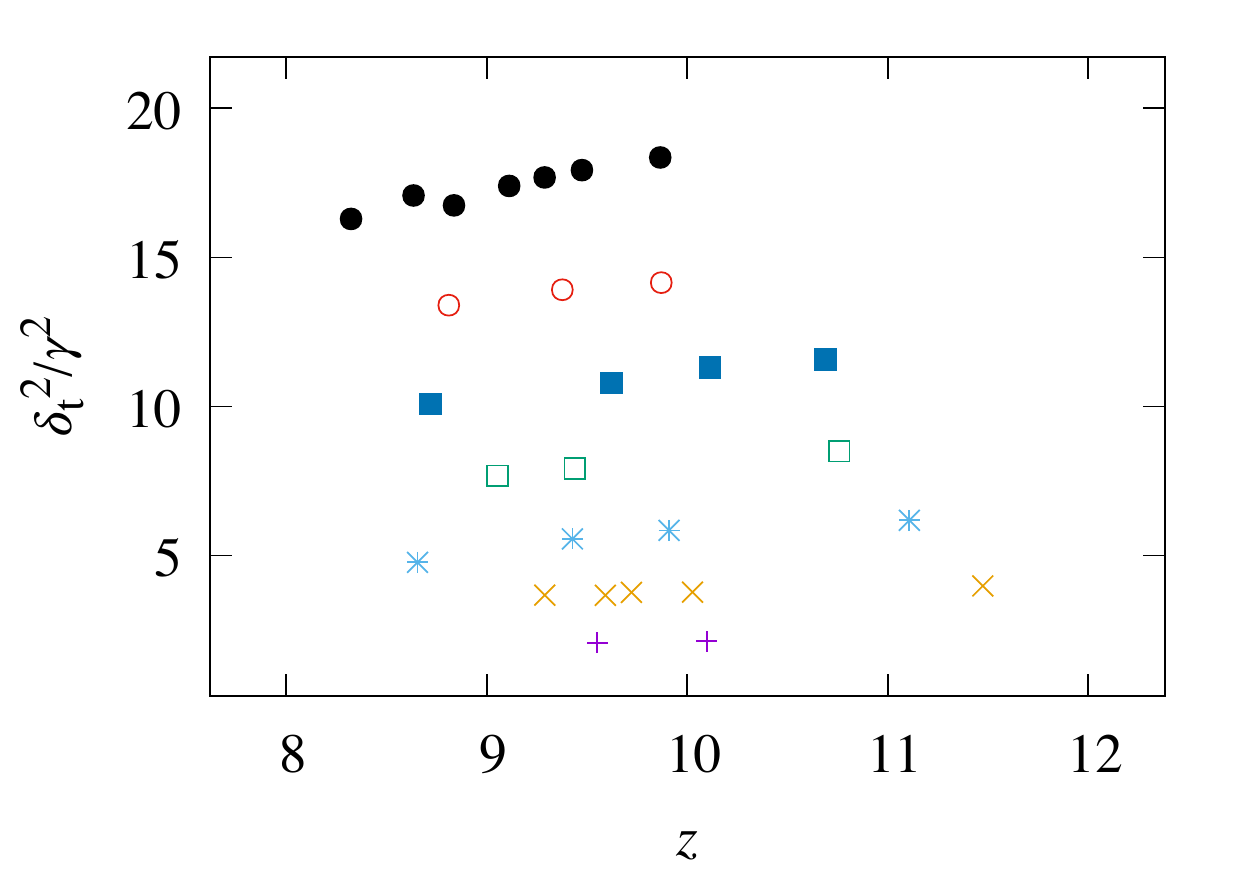}
  \includegraphics[width=0.32\textwidth]{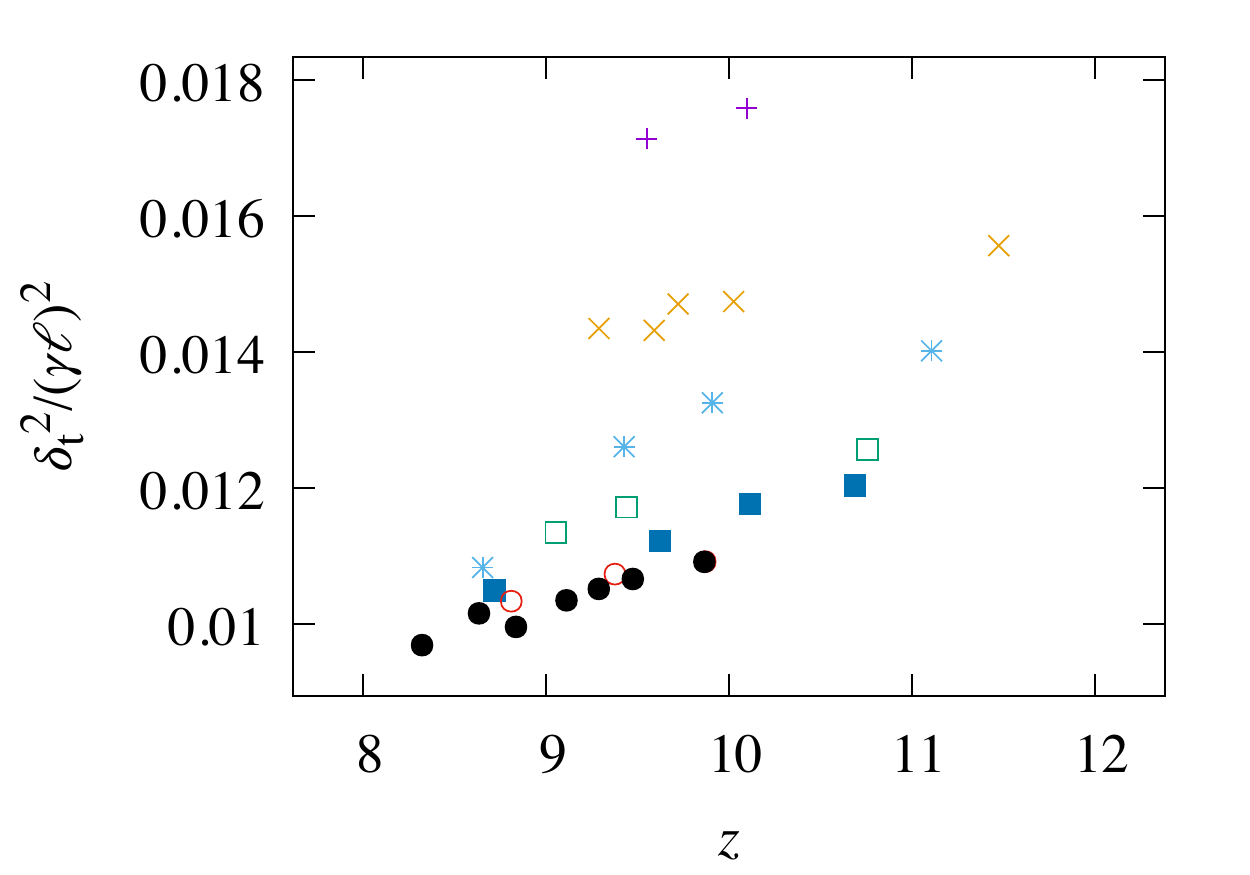}
  \includegraphics[width=0.32\textwidth]{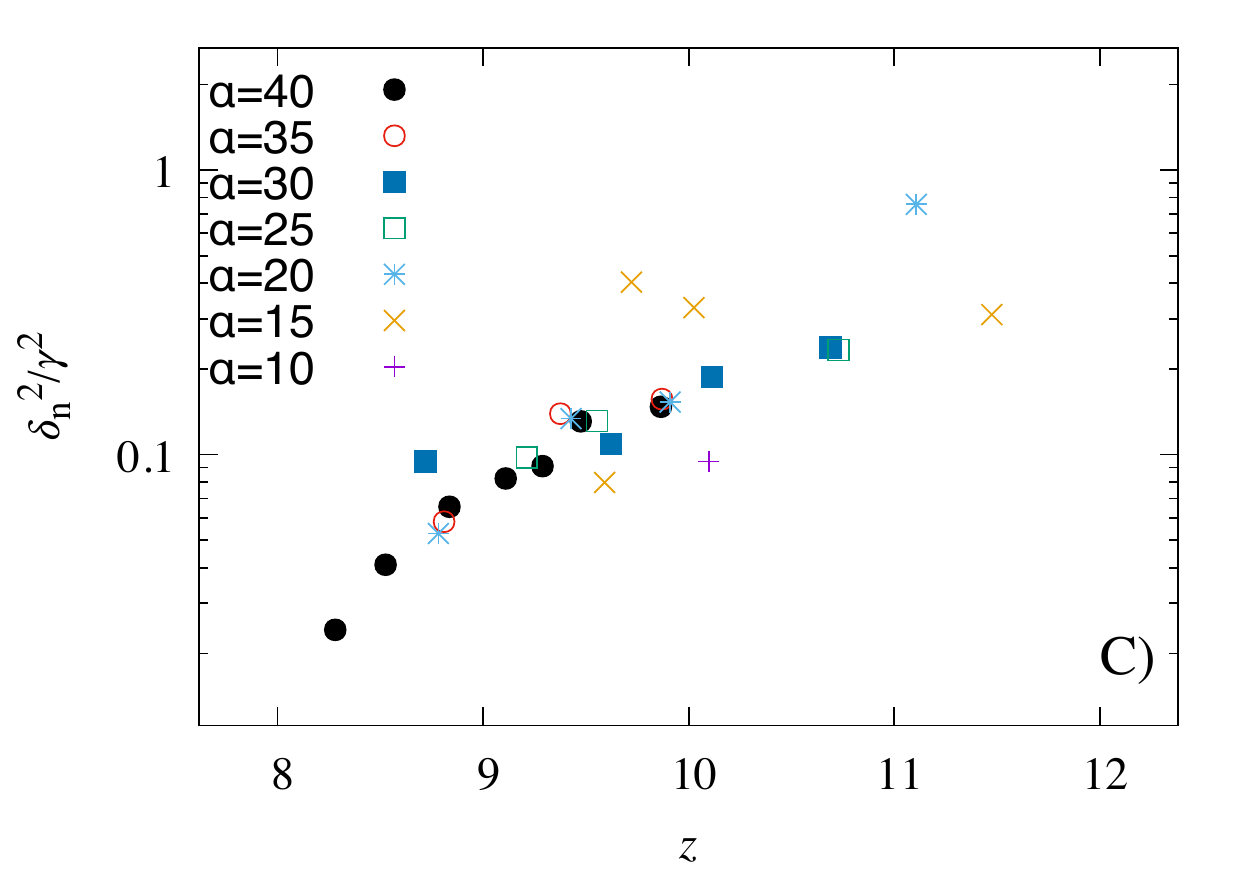}
  \caption{A) Mean-square tangential overlap $(\delta_t/\gamma)^2$
    vs. contacts $z$ for infinite friction $\mu\to\infty$ and various
    aspect ratios. B) same data, now rescaled by $\gamma\ell$; C)
    Mean-square normal overlap $(\delta_n/\gamma)^2$.}
  \label{fig:etang_z}
\end{figure*}

Comparing with the frictionless limit $\mu=0$ (see datapoints at
nearly zero, or Fig.~6 in Ref.~\cite{heussinger20:_packin}), the
infinite-friction modulus is two to three orders of magnitude
larger. Similarly large differences between frictional and
frictionless moduli have been observed in simulations of non-bonded
fibrous materials~\cite{PhysRevE.80.016115}.

Also the $\alpha$-dependence is different. The zero-friction modulus
$g_0$ vanishes in the long-SC limit, as $g_0\sim \alpha^{-2}$
\cite{heussinger20:_packin}. On the other hand, $g_\infty$ shows
hardly any dependence on $\alpha$ (Fig.~\ref{fig:modulus_z}), at least
in the large $\alpha$ limit.

SC length $\ell=\alpha d$ may enter the stress,
Eq.~(\ref{eq:stress_def}), and thus the modulus in various ways. The
first contribution comes from stress being an
energy-\emph{density}. The normalization with density $N/V$ can be
written as $N/V=\phi/V_{\rm sc}$ and, using Eqs.~(\ref{eq:v_sc}) and
~(\ref{eq:phi_rcp}), as $(d/\ell)/(d^2\ell) \sim \ell^{-2}$. This
alone would explain the result $g_0\sim \ell^{-2}$ of the frictionless
system.

However, additional $\ell$-dependence may come from the force-position
correlator part of the stress $\langle f_xR_y\rangle$. One
contribution is the force, or the magnitude of the overlaps $\delta$
and $\boldsymbol{\xi}^t$, and their change with strain, see
Eq.~(\ref{eq:contactforce_definition}). The normal overlaps $\delta$
are the only contribution in the frictionless system, and give rise to
$g_0$. In the frictional system also the tangential overlaps
$\boldsymbol{\xi}^t$ are present. They measure how much the positions
of contacts move on the surface of the SCs, in other words how
strongly SCs are sliding relative to each other. Such sliding motion
gives rise to frictional forces and thus to $g_\infty$.

In Fig.~\ref{fig:etang_z}A and B we display the mean-square tangential
overlap $\delta_t^2\equiv \langle
\boldsymbol{\xi}^t\cdot\boldsymbol{\xi}^t\rangle$. We find
$\delta_t\propto \gamma$, such that frictional contacts move at
constant velocity $v_t=\delta_t/t \propto \dot\gamma$. This velocity,
at least for large $\alpha$, is $\propto \ell$ as
Fig.~\ref{fig:etang_z}B shows. Thus, $\delta_t \propto \ell\gamma$
($v_t\propto \ell\dot\gamma$) and tangential displacements of contacts
per unit of strain grow with the length $\ell$ of the SC. The
displacements are, in particular, independent of the SC diameter $d$,
which is the alternative length-scale that might show up. On the other
hand, the normal overlaps $\delta_n^2 = \langle \delta^2\rangle$ are
independent of SC length, as panel C shows. They thus scale as
$\delta_n\propto d\gamma$.

Finally, there is also an $\ell$-dependence in the position part of
the correlator, which is the center-of-mass distance between the two
overlapping SCs (see Eq.~(\ref{eq:Rij}))
\begin{eqnarray}
  \mathbf{R}_{ij} &=& r_{ij}\mathbf{\hat e}_{ij}-\mathbf{\hat n}_is_i +
  \mathbf{\hat n}_js_j\,,
\end{eqnarray}
where the $\ell$-dependency is carried by the arclength parameters
$s_i$ and $s_j$.

In the case of pressure ($p\sim \langle \mathbf{f}\cdot\mathbf{R}
\rangle$) it is easy to see that this latter part does not contribute
in the frictionless scenario. There the force is normal to the
surface, $\mathbf{f}\parallel \mathbf{\hat e}$, and thus perpendicular
to the long axis given by $\mathbf{\hat n}$. The $s_i$, $s_j$ terms
therefore drop out. The same result is expected for the shear stress,
as long as one assumes statistical independence between the
orientation of the SC, $\mathbf{\hat n}$, and the orientation of the
contact with its neighbors. In the frictional scenario, on the other
hand, the force is tangential to the surface, thus the $s$-terms
survive.

\xx{Taken together all these dependencies one expects $g_0\sim
  z(\phi/d^2\ell)k_nd^2\sim (k_n/d)(d/\ell)^{2}$
  (Ref.\cite{heussinger20:_packin}) and $g_\infty\sim
  z(\phi/d^2\ell)k_t\ell^2\sim (k_t/d)$ (Fig.~\ref{fig:modulus_z}).}
\begin{figure}[ht]
  \includegraphics[width=0.4\textwidth]{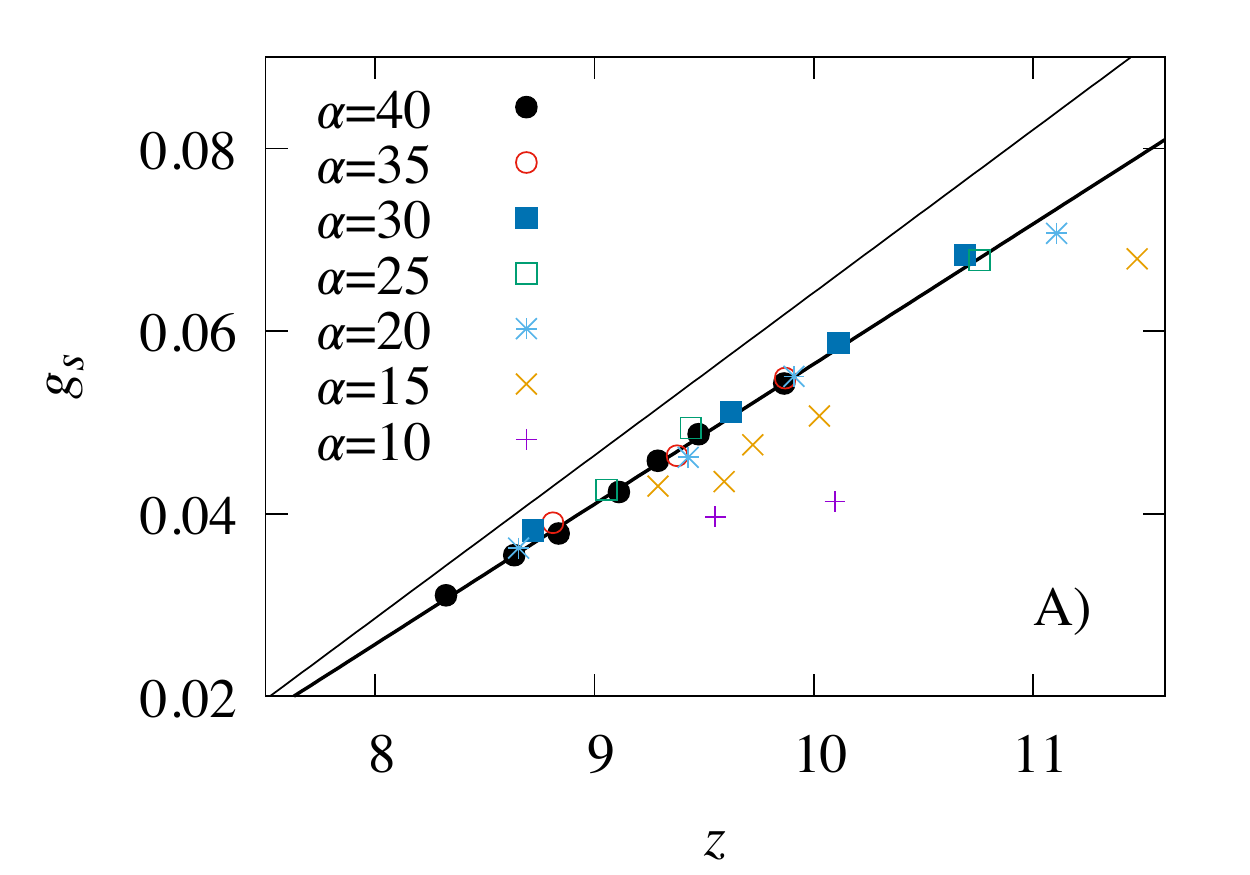}
  \caption{A) Shear modulus of the side contacts $g_s$ vs. $z$ for
    infinite friction $\mu\to\infty$ and various aspect ratios. Thin
    solid line full shear modulus (from Fig.~\ref{fig:modulus_z}) for
    the longest SC with $\alpha=40$. Thick solid line fit to the data
    with $\alpha=40$.  }
  \label{fig:side}
\end{figure}

\xx{Looking at Fig.~\ref{fig:modulus_z} one may also get the
  impression that the modulus slowly decreases with SC length. In
  fact, one can approximately collapse the data assuming
  $g_\infty\sim\alpha^{-1/2}$.  However, this apparent scaling is an
  artifact of the limited range of available $\alpha$. To justify this
  claim, we consider only the side contacts in the calculations of the
  modulus. This modulus $g_s$ is displayed in Fig.~\ref{fig:side}. In
  the limit $\alpha\to\infty$ the side contacts make the only
  contribution to the modulus as no end contacts occur. One clearly
  sees that this contribution is independent of $\alpha$, at least for
  the longest spherocylinders. Thus, we can safely assume that this
  value $g_s$, which is only slightly smaller than the full modulus
  $g_\infty$ (thin line, taken from Fig.~\ref{fig:modulus_z}),
  represents the finite full modulus in the $\alpha\to\infty$
  limit. In consequence, the scaling with $\alpha^{-0.5}$ cannot be
  true.}

\begin{figure}[ht]
  \includegraphics[width=0.4\textwidth]{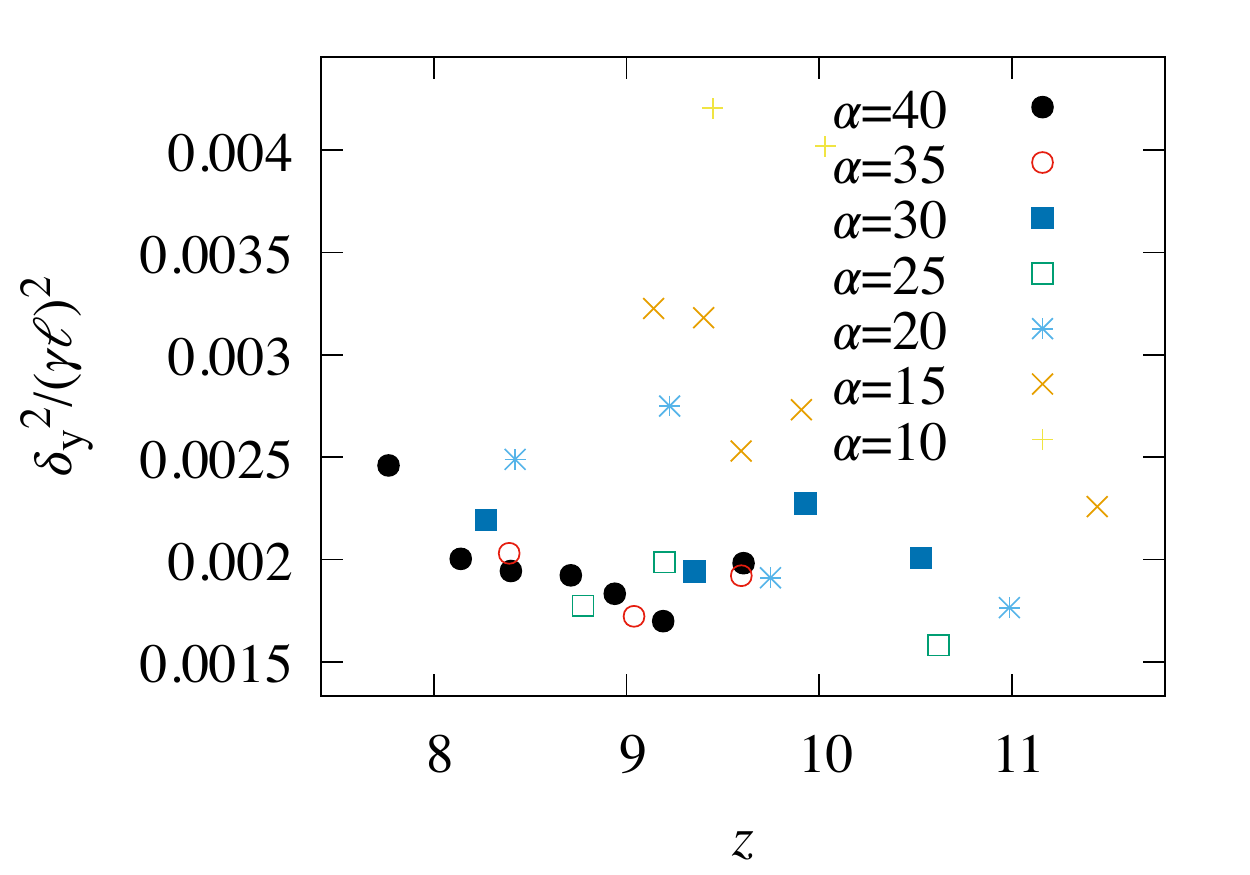}
  \caption{\xx{Fluctuations of SC center-of-mass displacement in
    gradient direction $(\delta_{y}/\gamma)^2$ vs. $z$.}}
  \label{fig:nonaffine}
\end{figure}

\xx{Finally, in Fig.~\ref{fig:nonaffine} we also consider the motion
  of particles (as compared to the motion of contacts as given by
  $\delta_t$). In response to the imposed strain $\gamma$, particles
  on average move in shear direction ($x$) by an amount $\gamma Y$,
  depending on the coordinate of the particle in the gradient
  direction ($y$). In gradient or vorticity direction there is no
  average motion. But fluctuations are present. In the figure we
  display the magnitude of fluctuating motion in gradient direction,
  $\delta_y(\gamma)^2 = \sum_i (Y_i(\gamma)-Y_i(0))^2/N$. At the small
  strains studied we find $\delta_y(\gamma)^2 \propto \gamma^2$. A
  different, e.g. diffusive, behavior may only be expected at much
  larger strains.}  Fluctuations in vorticity direction are of roughly
the same magnitude. As the figure shows, the fluctuations of the
motion of particles, similar to the displacements of contacts, scales
with $\ell$. Noteworthy, also the magnitude of both is quite similar,
$\delta_y\approx 0.05\gamma\ell$, to be compared with contact
displacements $\delta_t \approx 0.1\gamma\ell$.

\subsection{Finite friction coefficient $\mu<\infty$}

\begin{figure*}[ht]
  \includegraphics[width=0.32\textwidth]{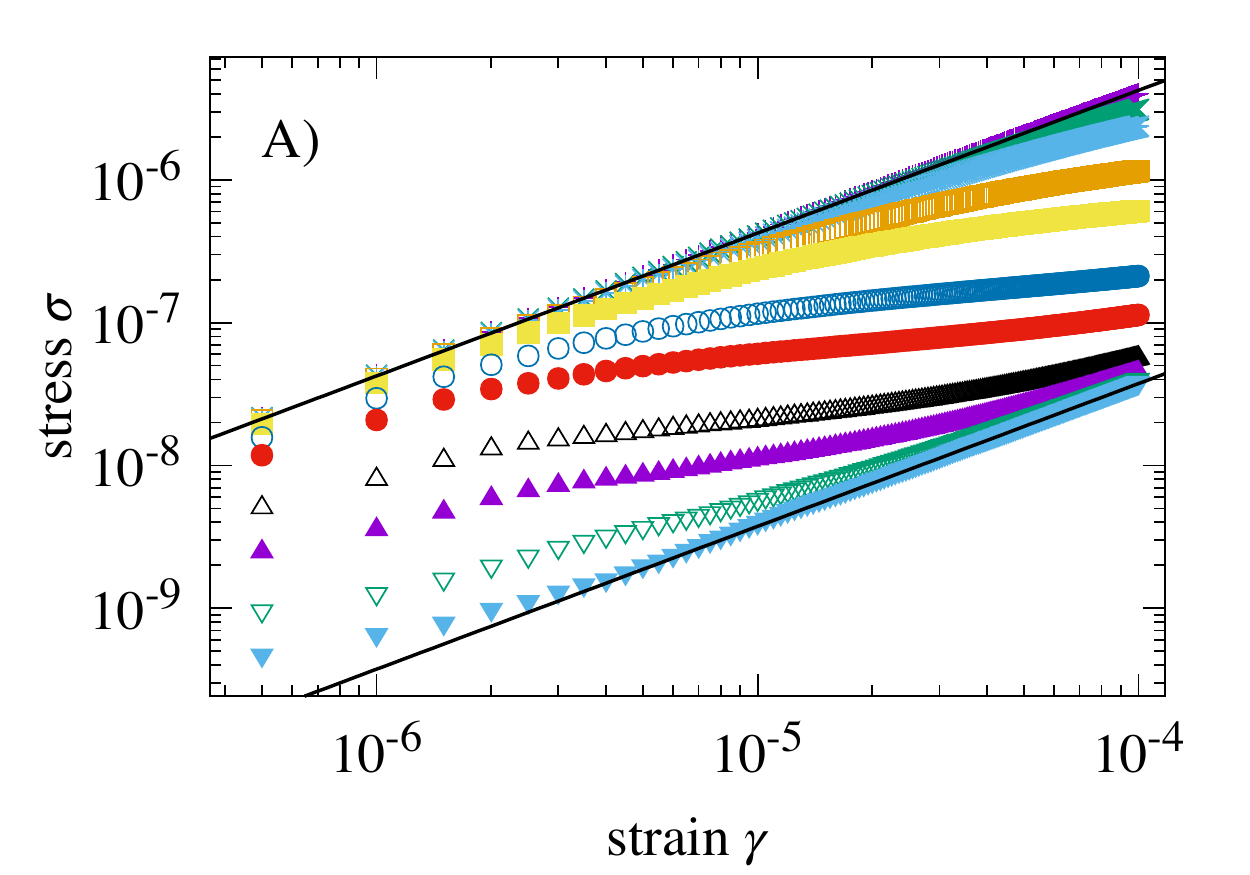}
  \includegraphics[width=0.32\textwidth]{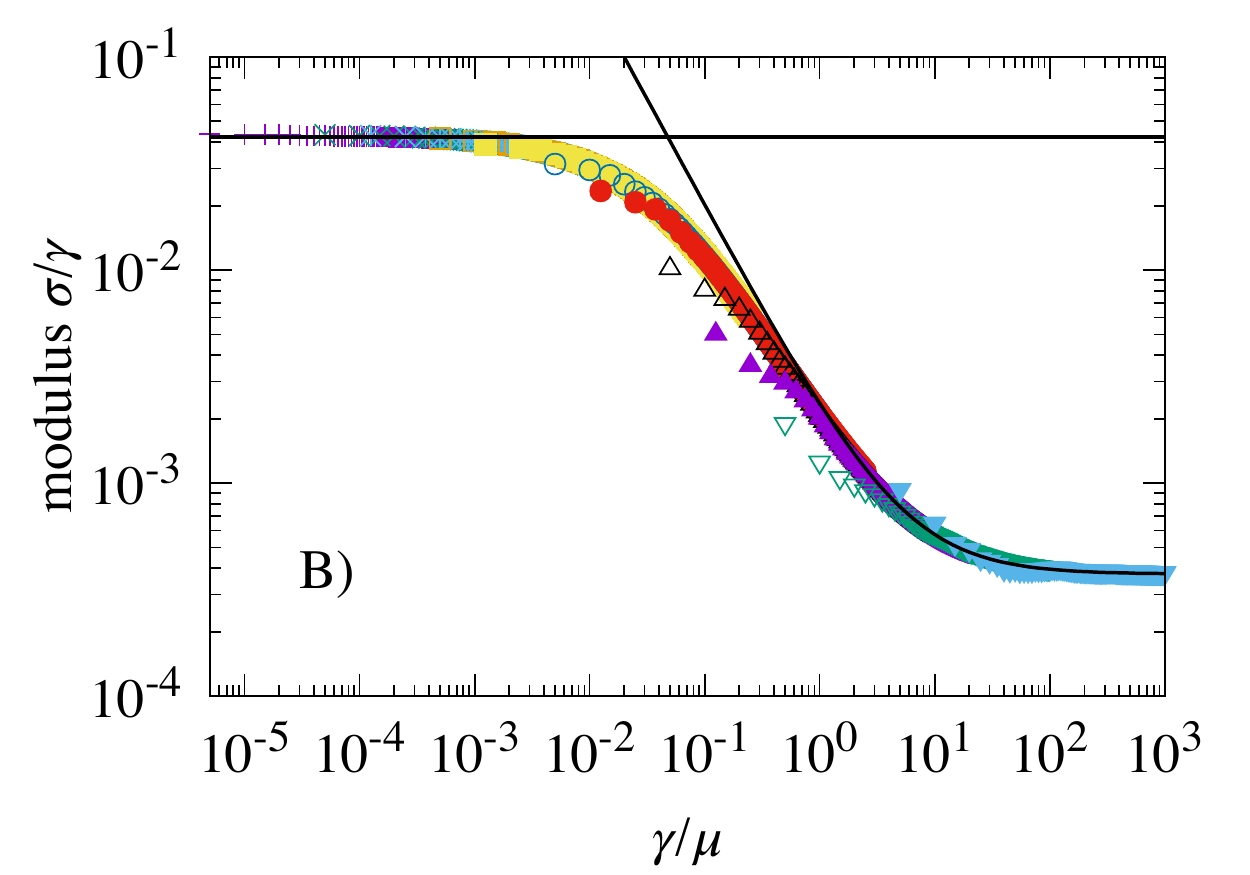}
  \includegraphics[width=0.32\textwidth]{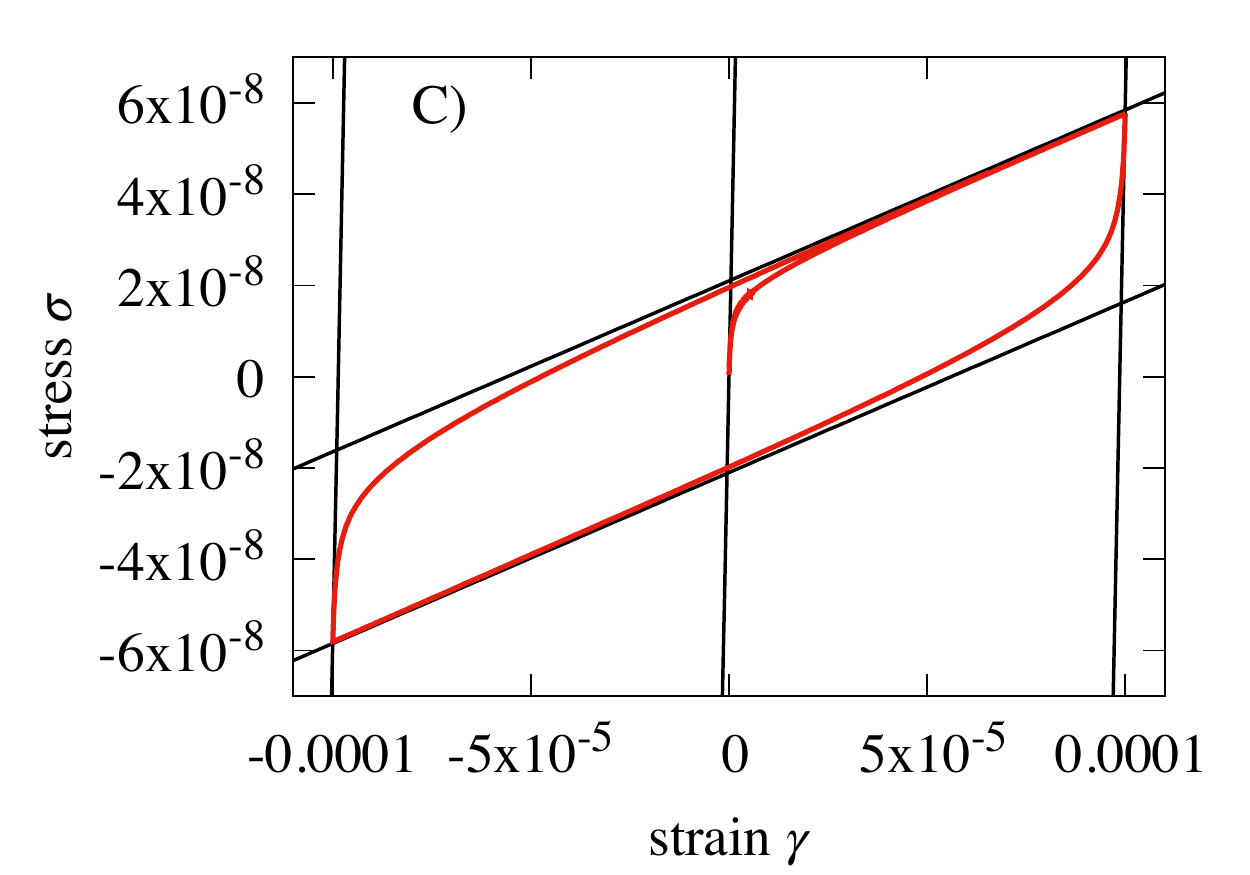}
  \caption{A) Stress vs. strain for various $\mu=1\ldots 10^{-7}$
    (from top to bottom) and $\alpha=40$, $\phi=0.11$. Lines represent
    limits of infinite (top) and zero (bottom) friction coefficient,
    respectively. B) Modulus $\sigma/\gamma$ vs. rescaled strain
    $\gamma/\mu$ for the same data. Solid lines: infinite-friction
    modulus and yield-stress model for zero-friction
    $\sigma=g_0\gamma+g_\infty\gamma_c$. Vertical line: crossover
    value $\gamma/\mu=0.05$ as derived in the text C) Strain cycle
    with maximum strain $\gamma_0=10^{-4}$ and $\mu=10^{-5}$. Lines
    represent limits of infinite and zero friction coefficient;
    shifted appropriately.}
  \label{fig:stress_varmu}
\end{figure*}

Infinite friction coefficient is rather unrealistic for any real
material. Here, we report results for varying friction coefficient,
spanning the range from nearly frictionless ($\mu\approx0$) to the
infinite-friction scenario of the first
section. Figure~\ref{fig:stress_varmu} presents the stress-strain
relation for different $\mu$ of one particular packing with
$\alpha=40$.
For small strain the stress follows the infinite-$\mu$ line
$\sigma = g_\infty\gamma$, then saturates in a quasi-plateau before
approaching the frictionless response $\sigma = g_0\gamma$. The strain
$\gamma_c$ where this change in behavior occurs depends on $\mu$.

At strains $\gamma<\gamma_c$, the external work $V\sigma\dot\gamma$ is
converted into ``potential'' energy of the tangential spring $k_t$:
$dE_{\rm pot}/dt\sim Nzk_t\ell^2\gamma\dot\gamma$. The stress here is
$\sigma = g_\infty\gamma$ with the infinite-friction modulus
$g_\infty\sim (\phi z/\ell)k_t\ell^2$, the prefactor assuring an
$\ell$-independent modulus, as discussed above.

The crossover scale to the quasi-plateau can be expected to depend on
the saturation of the Coulomb inequality, Eq.~(\ref{eq:coulomb}), as
equality, $f_t=\mu f_n$. At this point contacts start to slide, which
enters work-energy balance as additional dissipative term, $\Gamma_t
\sim Nz\mu f_nv_t$.

With the normal force set by pressure, $f_n\sim p\ell/z\phi$, we find
for the crossover strain
\begin{eqnarray}\label{eq:gamma_c}
\gamma_c \sim \frac{\mu p}{k_tz\phi}\,.
\end{eqnarray}
Indeed, all data collapse to a unique scaling form when plotted vs. a
rescaled strain $\gamma/\mu$
(Fig.~\ref{fig:stress_varmu}B). Respecting all the numerical factors
for this set of data, Eq.~(\ref{eq:gamma_c}) gives a crossover strain
$\gamma_c=0.05\mu$ (indicated by the vertical line in panel B).

For strains $\gamma>\gamma_c$ the relevant potential energy is that of
the normal overlaps, $dE_{\rm pot}/dt\sim Nzk_n\gamma\dot\gamma$, and
the stress becomes $\sigma \sim \sigma_y+g_0\gamma$ with the
zero-friction modulus $g_0\sim (\phi z/\ell)k_nd^2$ and the
``yield-stress'' $\sigma_y\sim g_\infty\gamma_c \sim \mu p\ell$. Thus,
at large strains the frictional forces contribute a yield stress
$\sigma_y\sim g_\infty\gamma_c\sim g_\infty\mu$ to the frictionless
response.

At zero strain the modulus is expected to reach the $\mu$-independent
value $g_\infty$. Deviations from this expectation and a lack of
scaling are visible in Fig.~\ref{fig:stress_varmu}B for some data
sets, which seem to level off at lower values. This, however, is an
artefact from a too large deformation rate $\dot\gamma=10^{-7}$. We
have checked that by reducing the strainrate to $10^{-8}$ no
deviations from scaling occur within the range of strains studied.

To wrap up: long spherocylinders respond to strain primarily via
sliding. Contacts are displaced in surface-tangential direction by
amounts, $\delta_t\sim \gamma\ell$. As a consequence, frictional
forces $f_t=k_t\delta_t$ increase and dominate the elastic
modulus. Non-frictional forces $f_n$, directed normal to the SC
surface, increase much slower and constitute only a small part of the
total modulus. At strains $\gamma_c\propto\mu$ the Coulomb threshold
of the contacts is reached and frictional forces cannot increase
further. The stress first reaches a plateau before, eventually, the
normal forces start to dominate the response. From this point the
packing behaves as if it were frictionless with a yield stress (the
plateau) from the frictional forces $\sigma_y\sim g_\infty\mu$. Still,
energy is dissipated because contacts are sliding. This gives rise to
hysteresis in oscillatory sweeps. To probe this we perform oscillatory
strains $\gamma(t)=\gamma_0\sin(\omega t)$
(Fig.~\ref{fig:stress_varmu}C) with variable maximal strain $\gamma_0$
and frequency $\omega$ chosen such that the product $\gamma_0\omega$
matches the strainrates used up to now. Friction governs the response
(highlighted by the steep lines $\sigma=g_\infty\gamma$) in the
startup, as well just after strain reversal. After reversal the
frictional sliding stops and contacts stick. The tangential springs
relax and load in the opposite direction. Once the sliding limit in
this direction is reached, a crossover to frictionless response
(shallow lines $\sigma=g_0\gamma$) is
observed again.

\subsection{Viscous dissipation}

Dissipation is due to sliding friction of the contacts. Technically,
this is due to the application of the Coulomb inequality,
Eq.~(\ref{eq:coulomb}), which effectively rescales $\xi^t$ to keep
tangential forces from increasing beyond what is allowed by the
inequality. This leads to a loss of energy. This mechanism is
strainrate independent and is therefore called ``plastic
dissipation''. The rescaling also implies that $c_t$, i.e. the viscous
dissipation in the tangetial motion is not relevant; it is the total
tangential force, including the viscous component, that is limited
according to Coulomb inequality.
To assess the influence of viscous dissipation on the response we
therefore study the frictionless limit $\mu=0$ in combination with
different values of $c_t$, which governs the strength of viscous
dissipation due to relative sliding motions. The tangential force,
Eq.~(\ref{eq:contactforce_definition}), reduces to
$\mathbf{f}_t=-c_t\mathbf{v}_t$.

\begin{figure}[ht]
  \includegraphics[width=0.4\textwidth]{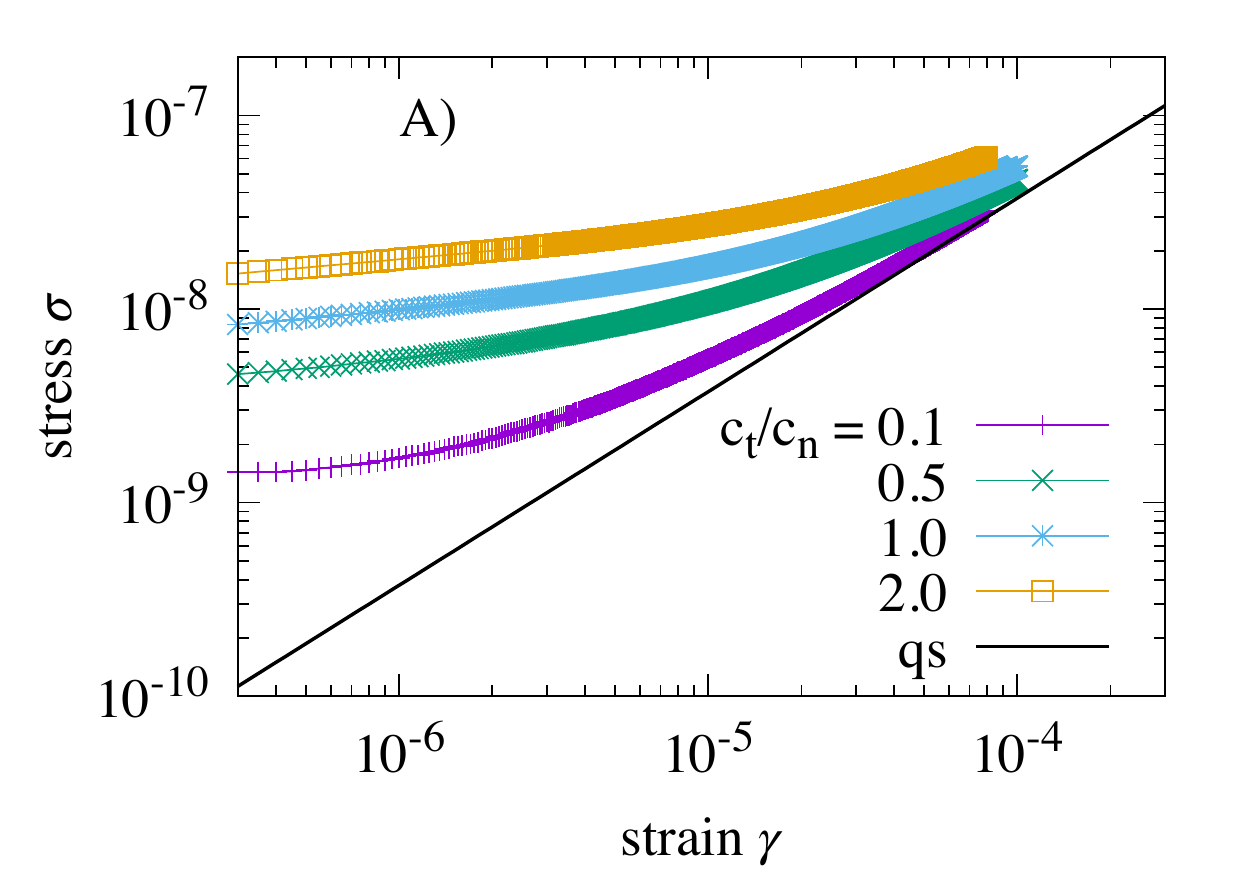}
  \includegraphics[width=0.4\textwidth]{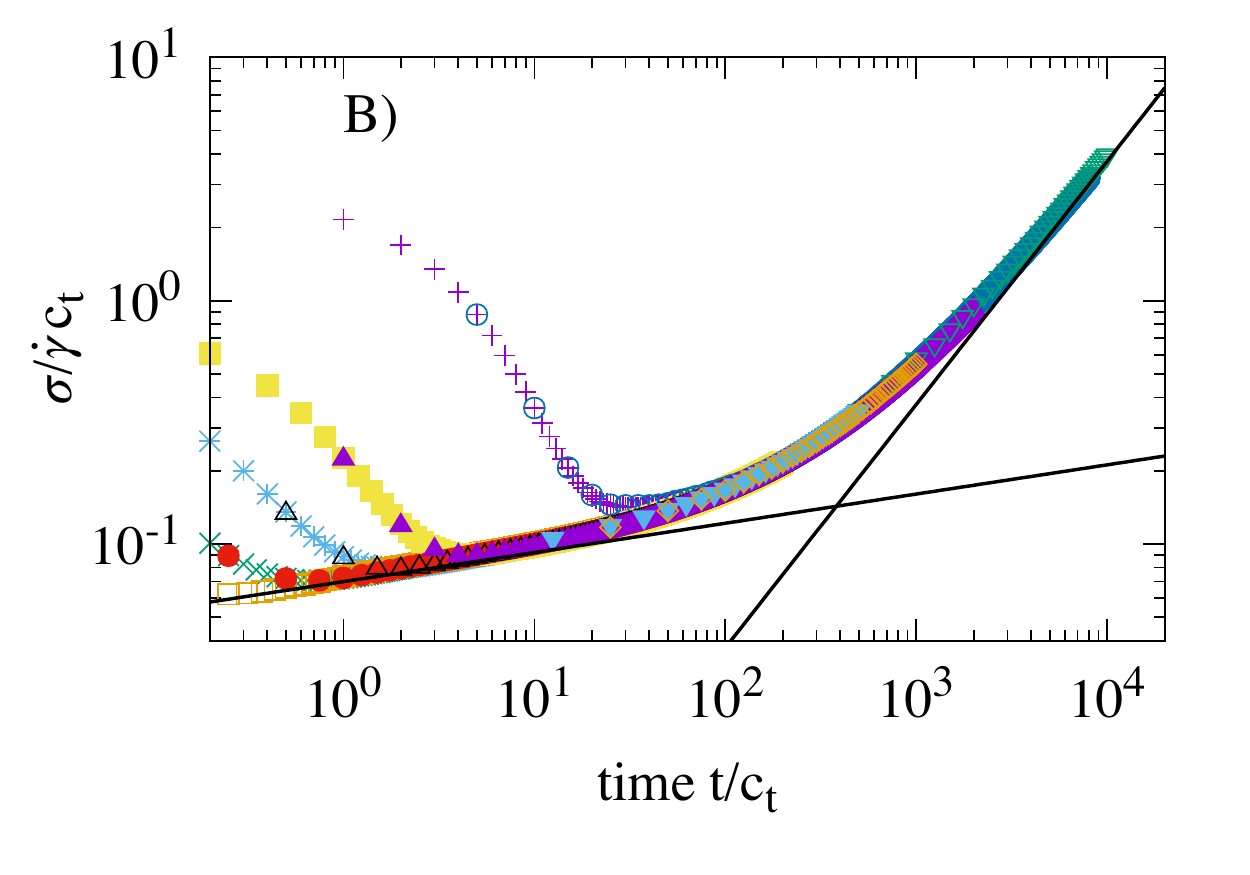}
  \caption{A) Stress vs. strain for various $c_t/c_n=0.1\ldots 4.0$,
    $c_n=0.5$.  and $\alpha=40$, $\phi=0.11$, $\mu=0$,
    $\dot\gamma=10^{-7}$. Line represents quasi-static results, see
    Fig.~\ref{fig:stress_varmu} B) Rescaled stress
    $\sigma/(\dot\gamma c_t)$ vs. time $t/c_t$ for the same data, as
    well as different strainrates,
    $\dot\gamma=10^{-6},10^{-8}$. Dotted line $\sim x^\epsilon$, $\sim
    x$.}
  \label{fig:stress_varct}
\end{figure}

Fig.~\ref{fig:stress_varct} compares the stress-strain relations for
one particular packing sheared at various strainrates and with
different values of $c_t$. For comparison the quasistatic result, for
which the parameter $c_t$ is irrelevant, is also displayed.

The data indicate the scaling form $\sigma = c_t\dot\gamma F(t/c_t)$,
with $F(x\to\infty)\to x$ or $F(x\to0)\to x^\epsilon$ and small
exponent $\epsilon\approx 0.13$. An $\epsilon=0$ would give $\sigma =
c_t\dot\gamma$ in the initial regime, which indicates dissipation via
tangential viscous forces when contacts are sliding with velocities
set by the strainrate $v_t\sim \ell\dot\gamma$. In more detail, the
energy-work balance has three terms
\begin{eqnarray}
Nz k_n\gamma\dot\gamma = V\sigma\dot\gamma - Nzc_t\langle v_t^2\rangle\,.
\end{eqnarray}
The terms are the time-rate of change of the potential energy, the
external work and the dissipation now via viscous forces. At early
times one balances stress with dissipation to obtain $\sigma \sim
(\phi z/\ell)c_t\dot\gamma\ell^2$, i.e. a time-independent
constant. This is the regime, where, in our data, we still see a weak
time-dependence, governed by the exponent $\epsilon$.  For the
velocities this would imply a time-dependence
$v_t\sim (t/c_t)^{\epsilon/2}\ell\dot\gamma$.
At long times the usual elastic regime $\sigma = g_0\gamma$ sets
in. The crossover time-scale is $\tau \sim z^2c_t/g_0$
At very short times $t\sim 1$, the scaling with $(t/c_t)^\epsilon$
breaks down. This corresponds to the elastic collision time-scale,
i.e. the very first collision when initial conditions are still
important. In the figure this is visible as the hump at small times.

\section{Discussion}

We have dealt with assemblies of long spherocylinders (SC). Because of
the large excluded volume of such high-aspect ratio particles, packings
are of very low volume-fraction that decreases with particle length as
$\phi\propto \ell^{-1}$.

The key question we posed was in how far frictional forces modify or
dominate the elastic response at small strains. Starting point was the
assumption that steric hindrance is the key factor to restrict
particle motion. Free motion is then only possible along the long axis
of the cylinder. This motion induces sliding of the contacts on the
surface of the cylinders and gives rise to large frictional forces.

We have argued that frictional contact forces act similar to forces
from an elastic bond, at least as long as the Coulomb threshold is not
yet reached. In this analogy the motion of a contact on the surface is
comparable to the extension of a permanent elastic bond (here
described via a spring constant $k_t$). Such permanent bonds
frequently occur in biological systems, e.g. the cytoskeleton, where
the long filamentous or rod-like polymers are chemically, or
electro-statically bonded via crosslinking proteins
\cite{bausch06,alb94}. Depending on the stiffness of the bonding and
the polymers' intrinsic elasticity the described mechanism of contact
sliding/bond extension might also lead to secondary (stretching,
bending) deformations in the
polymers~\cite{C1SM05022B,PhysRevE.93.062502,heussingerPRE2007}. Note,
that the spherocylindrical particles used here do not have these
degrees of freedom; they are modelled as cylinders with a straight
backbone that may not change length. Biopolymers in cytoskeletal
systems usually have very high aspect-ratios with diameters in the
nanometer range and lengths exceeding the $\mu$m scale. For rod-like
microtubules, however, with a diameter of $d\approx 25$nm and lengths
$\ell\approx \mu$m our simulations with $\alpha=40$ are within
physically reasonable values also for these particles.

We have determined the shear-induced tangential motion $\delta_t$ of
contacts on the surface of particles (the extension of the ``bonds'')
in a variety of packings with spherocylinders of different lengths
$\ell$. Interestingly, $\delta_t\propto \ell$ and thus increases with
the length of the particles. This can be understood by assuming the
packing to respond affinely to an imposed shear deformation
$\gamma$. From the properties of an affine map, the distance between
the center-of-mass of two SCs should change in proportion to their
distance. Overlapping SCs have distances on the order of their length
$\ell$, such that also the change in distance is,
$\Delta\ell\propto\gamma\ell$. As the SCs themselves do not change
length, the motion of the center-of-mass gives rise to relative
sliding motion of the contacts of exactly this order of magnitude,
$\delta_t\propto \gamma\ell$. While we have seen
(Fig.~\ref{fig:nonaffine}) that the actual motion of the SCs also
includes a substantial fluctuating component, this additional
component also scales with $\ell$. Thus, the overall scaling
$\delta_t\propto\ell$ is not affected, albeit the prefactor is
changed.

The shear modulus itself has a finite (non-zero) limit for large
$\ell$. We have shown that this results from the combined effect of
increasing $\delta_t$ and decreasing overall density $\phi$. On the
other hand, the modulus in the absence of friction is orders of
magnitudes smaller and vanishes as $g_0\propto \ell^{-2}$. Contact
motion responsible for forces in this frictionless limit is normal to
the surface of the particles, $\delta_n$. Without friction, tangential
motion $\delta_t$ does not build-up forces. We find that $\delta_n$ is
much smaller than $\delta_t$ and does not scale with the length of the
SCs but their diameter, $\delta_n\sim d$.

As the strain increases frictional forces reach the limit set by the
Coulomb inequality Eq.~(\ref{eq:coulomb}). At this point the packing
starts to ``shear-thin'' and the shear modulus decreases to its
frictionless value $g_0$. This generally happens at strains
$\gamma_c\sim \mu p /k_tz\phi\sim \mu/\ell$ the latter represents the
limiting behavior for long SCs.
At these strains, hysteresis is observed in oscillatory sweeps
$\gamma(t) = \gamma_0\sin(\omega t)$. This highlights the presence of
energy dissipation in the sliding contacts and evidences a transition
from static ($\gamma<\gamma_c$) to dynamic friction
($\gamma>\gamma_c$).

Finally, we also consider energy dissipation via viscous forces,
embodied in the parameter $c_t$. With the assumption of
time-independent velocities $v_t\sim \ell\dot\gamma$ one expects a
time-independent stress $\sigma \sim c_t\dot\gamma$ at small
strains. Rather we obtain $\sigma \sim c_t\dot\gamma(t/c_t)^\epsilon$
with a small exponent $\epsilon\approx 0.1$ that embodies the
time-dependence of the stress. For the contact velocities this would
imply $v\sim (t/c_t)^{\epsilon/2}\ell\dot\gamma$, alternatively for
the contact displacements $\delta_t^2 \sim t^{2+\epsilon}$. The origin
of such a behavior is currently unclear.

\begin{acknowledgments}

  We acknowledge financial support by the German Science Foundation
  (Deutsche Forschungsgemeinschaft) via the Heisenberg program (CH:
  HE-6322/2).
  
\end{acknowledgments}

\end{document}